\documentclass[aps,prl,showpacs,twocolumn,superscriptaddress]{revtex4}

\usepackage{graphicx}
\usepackage{bm}
\begin{document}
\title{Influence of domain wall interactions on nanosecond switching in magnetic tunnel junctions}

\author{J.~Vogel}
\affiliation{Laboratoire Louis N\'{e}el, CNRS, 25 avenue des
Martyrs, B.P. 166, F-38042 Grenoble Cedex 9, France}
\author{W.~Kuch}
\altaffiliation[Present address: ]{Institut f\"{u}r
Experimentalphysik, Freie Universit\"{a}t Berlin, Arnimallee 14,
D-14195 Berlin, Germany.} \affiliation{Max-Planck-Institut f\"{u}r
Mikrostrukturphysik, Weinberg 2, D-06120 Halle, Germany}
\author{R.~Hertel}
\affiliation{Institute of Solid State Research (IFF), Research Center
J\"{u}lich, D-52425 J\"{u}lich, Germany}
\author{J.~Camarero}
\affiliation{Dpto. F\'{i}sica de la Materia Condensada, Universidad
Aut\'{o}noma de Madrid, E-28049 Madrid, Spain}
\author{K.~Fukumoto}
\altaffiliation[Present address: ]{Institut f\"{u}r
Experimentalphysik, Freie Universit\"{a}t Berlin, Arnimallee 14,
D-14195 Berlin, Germany.} \affiliation{Max-Planck-Institut f\"{u}r
Mikrostrukturphysik, Weinberg 2, D-06120 Halle, Germany}
\author{F.~Romanens}
\affiliation{Laboratoire Louis N\'{e}el, CNRS, 25 avenue des
Martyrs, B.P. 166, F-38042 Grenoble Cedex 9, France}
\author{S.~Pizzini}
\affiliation{Laboratoire Louis N\'{e}el, CNRS, 25 avenue des
Martyrs, B.P. 166, F-38042 Grenoble Cedex 9, France}
\author{M.~Bonfim}
\affiliation{Departamento de Engenharia El\'{e}trica, Universidade
do Paran\'{a}, CEP 81531-990, Curitiba, Brazil}
\author{F.~Petroff}
\affiliation{Unit\'{e} Mixte de Physique CNRS/Thales, Route
d\'{e}partementale 128, F-91767 Palaiseau, France, and
Universit\'{e} Paris-Sud, F-91405 Orsay, France}
\author{A.~Fontaine}
\affiliation{Laboratoire Louis N\'{e}el, CNRS, 25 avenue des
Martyrs, B.P. 166, F-38042 Grenoble Cedex 9, France}
\author{J.~Kirschner}
\affiliation{Max-Planck-Institut f\"{u}r Mikrostrukturphysik,
Weinberg 2, D-06120 Halle, Germany}


\begin{abstract}
We have obtained microscopic evidence of the influence of domain
wall stray fields on the nanosecond magnetization switching in
magnetic trilayer systems. The nucleation barrier initiating the
magnetic switching of the soft magnetic Fe$_{20}$Ni$_{80}$ layer in
magnetic tunnel junction-like FeNi/Al$_2$O$_3$/Co trilayers is
considerably lowered by stray fields generated by domain walls
present in the hard magnetic Co layer. This internal bias field can
significantly increase the local switching speed of the soft layer.
The effect is visualized using nanosecond time- and layer-resolved
magnetic domain imaging and confirmed by micromagnetic simulations.
\end{abstract}

\pacs{75.60.Jk, 75.60.Ch, 75.70.-i, 85.70.Kh, 07.85.Qe}

\maketitle

The active part of devices like spin-valves and magnetic tunnel
junctions, used in magnetic random access memories (MRAM), consist
of an ultrathin soft magnetic layer and a harder magnetic layer
separated by a non-magnetic spacer layer. These devices rely on the
fast switching of the magnetization of the soft layer for reading or
writing separate bits of information. Micromagnetic interactions
have a strong influence on this switching. Demagnetizing effects and
stray fields at the edges of nanosized magnetic structures can
influence the magnetic configuration and the magnetization reversal
of the soft magnetic layer, but interface roughness can also play a
role and induce a magnetostatic coupling with the underlying hard
magnetic layer. Much larger, but more localized magnetostatic
effects exist when a domain wall is present in the hard magnetic
layer \cite{fuller62,Hubertbook}. Direct evidence of the influence
of domain wall stray fields in one layer on the static domain
configuration of another layer has been obtained by Kuch \textit{et
al.} \cite{kuch03} on Co/Cu/Ni trilayers using x-ray photoelectron
emission microscopy (X-PEEM). Sch\"{a}fer \textit{et al.}
\cite{schaf02,schaf04} have used Kerr microscopy to show the effect
of stray fields of Bloch domain walls in a Fe whisker on the
magnetization of a thin Fe film through a MgO spacer. Similar
effects were recently also observed in systems with perpendicular
magnetization \cite{Wiebel05}. Thomas \textit{et al.}
\cite{thomas00} have observed that repeated motion of domain walls
in the soft magnetic layer of a soft magnetic/nonmagnetic/hard
magnetic trilayer can demagnetize the hard magnetic layer, even if
the coercive field of the hard layer is several times larger than
the field used for the reversal. In thin films with in-plane
uniaxial anisotropy the static coercivity is usually determined by
the field needed for domain nucleation. In FM/NM/FM trilayers the
stray field of a domain wall in the hard magnetic layer can locally
decrease this quasi-static nucleation field in the soft magnetic
layer \cite{fuller62}. Here, we show a first direct, real-time
observation of this effect by studying the magnetization reversal
dynamics of the soft FeNi layer in the presence of a domain wall in
the harder Co layer in Fe$_{20}$Ni$_{80}$/Al$_2$O$_3$/Co trilayers,
taking advantage of the layer selectivity of X-PEEM combined with
x-ray magnetic circular dichroism (XMCD-PEEM). Our micromagnetic
simulations show that the stray field of domain walls in the Co
layer locally tilts the magnetization of the Fe$_{20}$Ni$_{80}$
(FeNi in the following) layer in the direction perpendicular to the
easy axis, opposite to the direction of the magnetization in the
core of the Co domain wall. The Co domain wall stray field acts thus
locally as an effective transverse bias field \cite{choi01}. This
internal bias field decreases the nucleation barrier and can
considerably increase the local switching speed of the soft layer.

The measurements were performed on a
Fe$_{20}$Ni$_{80}$(4nm)/Al$_2$O$_3$(2.6nm)/Co(7nm) trilayer
deposited on Si(111) by RF sputtering. The Si substrate was miscut
by 6$^{\circ}$ along the (\=211) direction, which after heat
treatment \cite{Sussiau} leads to a step-bunched surface presenting
terraces in the shape of ellipses with an average length of about
1~$\mu$m and a width of about 40~nm, separated by 6~nm high steps.
Before depositing the trilayer, 3 nm of Co were deposited and then
oxidized to form a layer of CoO. This layer served to increase the
coercivity of the Co layer and allowed doing measurements with
magnetic pulses strong enough to saturate the FeNi layer without
changing the Co domain pattern. The alumina layer was obtained by
depositing pure Al followed by a glow discharge under a 10 Pa O$_2$
plasma. The topographic steps in the substrate are transferred to
the magnetic trilayers \cite{Montaigne,pennec04}, leading to an
in-plane uniaxial magnetic anisotropy with the easy axis along the
long axis of the terraces. The steps at the end of the terraces
induce a magnetostatic N\'{e}el orange-peel coupling
\cite{Neel62,encinas03,pennec04} between the two magnetic layers
through the spacer layer. In Fig.~1 we present magnetization loops
of the sample, obtained by longitudinal Kerr effect measurements,
for fields applied either along (easy axis) or perpendicular (hard
axis) to the steps. Minor loops for the permalloy layer show a shift
of about 1 mT with respect to zero field due to the orange-peel
coupling. The squareness of the loops indicates that in quasi-static
conditions the reversal takes place through the nucleation of one or
several reversed domains and a subsequent fast propagation of the
generated domain walls. The coercivity is determined by the
nucleation barrier.

\begin{figure}
\includegraphics*[bb= 216 475 377 594]{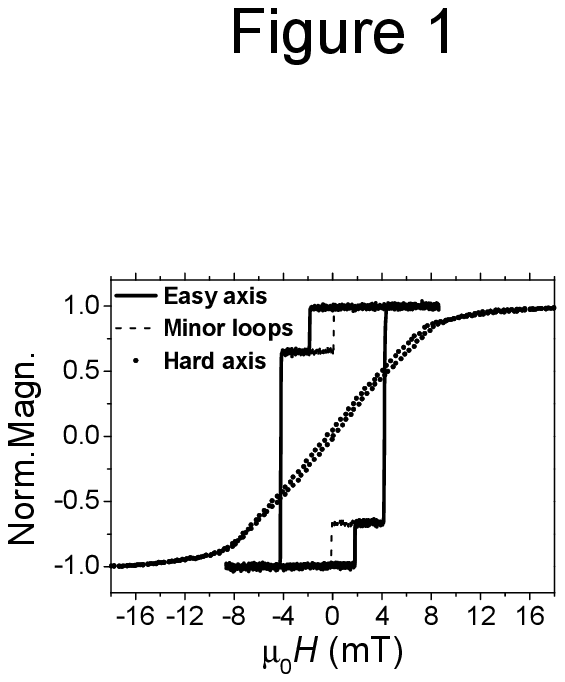}
\caption{Quasi-static hysteresis loops of the magnetic tunnel
junction-like trilayer obtained by longitudinal Kerr effect. Loops
obtained with the field applied along the easy (continuous line) and
hard (dotted line) magnetization axes are shown, as well as minor
loops for the FeNi layer taken along the easy axis (dashed line).}
\end{figure}

Time- and layer-resolved magnetic domain images were obtained
combining XMCD-PEEM and a stroboscopic acquisition mode
\cite{vogel03,kuch04,schneider04}. The used microscope was a
commercial Focus IS-PEEM and the experimental setup was identical to
the one described in previous publications \cite{kuch98}. Magnetic
pulses provided by a small coil mounted directly on the sample were
synchronized with the x-ray photon pulses \cite{vogel03}, with a
repetition rate of 625 or 312.5 kHz. The measurements were performed
on beamline UE56/2-PGM2 at the BESSY synchrotron in Berlin
(Germany). The FeNi domain structure was imaged by tuning the x-ray
energy to the Fe L$_3$ absorption edge (707 eV), while for the Co
layer the Co L$_3$ edge energy (778 eV) was used.

\begin{figure}
\includegraphics[width=6cm]{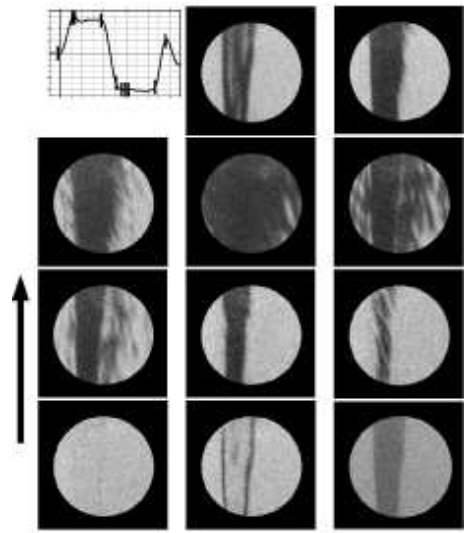}
\caption{Time and layer-resolved XMCD-PEEM images of the
magnetization state of the FeNi [(b)-(k)] and Co layers (l). The
field of view in these images is about 100 $\mu$m and the spatial
resolution 1 $\mu$m. The projection of the x-ray incidence direction
on the sample surface is pointing up in the images (parallel to the
arrow) and is parallel (anti-parallel) to the direction of the field
for positive (negative) pulses. The magnetization direction is in
the plane of the layers and points up (parallel to the arrow) for
black domains, and down for white domains. The FeNi images were
taken for delays between photon and magnetic pulses of -3, 11, 13,
37, 51, 56, 60, 62, 87 and 97 ns, as indicated in (a). The Co image
was taken for a delay of 60 ns.}
\end{figure}

In order to study the fast switching of the FeNi layer and the
influence of the Co domain walls thereon, we first induced a domain
structure in the Co layer using a 3 ms pulse with an amplitude of
about 10 mT. The resulting domain structure is shown in Fig.~2(l).
White and black regions correspond to domains with an in-plane
magnetization direction pointing parallel and anti-parallel to the
black arrow, respectively. Bipolar magnetic pulses with amplitude of
about 6 mT and a length of 40 ns for both positive and negative
pulses were then applied to the sample parallel to the easy
magnetization axis. Images recorded for the FeNi layer in pump-probe
mode, for different delays between photon and magnetic pulses, are
shown in Fig.~2(b)-2(k). The corresponding positions of the photon
pulses with respect to the magnetic pulses are given in Fig.~2(a).
Before the pulses [Fig.~2(b)] the domain structure in the FeNi layer
is strongly correlated to the one in the Co layer [Fig.~2(l)], due
to the rather strong orange-peel coupling that tends to align the
magnetization direction in the two magnetic layers. The small
overshoot at the end of the negative pulse is not sufficient to
completely align the magnetization in the FeNi layer above the black
Co domain with the underlying Co magnetization before the pulse.
When the field increases, propagation of the domain walls in the
FeNi layer takes place first [Fig.~2(c)]. At the maximum of the
positive pulse some newly nucleated reversed domains also become
visible [Fig.~2(d)]. At the end of the positive pulse, the FeNi
layer is almost saturated except for some remaining white domains at
the right and left bottom [Fig.~2(e)]. When the field is reversed,
starting from this nearly saturated state, new white domains appear
during the rising flank of the pulse [Fig.~2(f)] above the white
domains in the Co layer and above the Co domain walls [Fig.~2(f,g)].
Nucleation and propagation of domain walls is faster above the white
domains in the Co layer, due to the orange peel coupling leading to
a higher effective field (external plus coupling fields). On the
plateau of the negative pulse, the FeNi above the black Co domain
switches, initially by propagation of domain walls [Fig.~2(h)] and a
few nanoseconds later also by nucleation of some new white domains
[Fig.~2(i)]. At the end of the negative pulse [Fig.~2(j)], the FeNi
layer seems completely saturated in the white direction. Closer
inspection actually shows that faint grey lines are still present at
the position where domain walls are present in the Co layer. This
shows that the Co domain wall stray field is strong enough to
prevent saturation of the FeNi layer even for a field of 6 mT, which
is about three times the static coercivity for the FeNi layer. On
the other hand when the field direction is changed, preferential
nucleation of reversed domains takes place above the domain wall in
the Co layer, as seen in Fig.~2(k). This effect also takes place
going from e) to g) but it is less clear since nucleation centers
become visible only after some time, when they have expanded through
domain wall propagation to a size that is large enough to be visible
with our microscope. Nucleation processes in the FeNi layer may
occur on top of the Co domain walls, but at higher fields also in
the middle of existing Co domains. If the field is increased very
quickly, the corresponding critical fields are reached almost at the
same time, so that the difference between the two nucleation fields
is difficult to see (cf. 2(f,g)). In 2(k), the field was increased
more slowly and to lower values, making this difference in
nucleation field (and time) much more visible.

Our nanosecond time- and layer-resolved domain images thus reveal
the strong influence of domain walls in the Co layer on the
nanosecond reversal of the FeNi layer. We did several measurement
series on the same spot of the sample, with different domain
patterns induced in the Co layer. Preferential nucleation always
takes place above Co domain walls and is therefore not simply caused
by topographic features in the sample. Other samples, with different
miscuts and anisotropies showed the same qualitative results. This
clearly indicates that the Co domain wall stray field locally
decreases the barrier for nucleation of reversed domains in the
adjacent FeNi layer, therefore increasing the local switching speed.

\begin{figure}
\includegraphics[width=6cm]{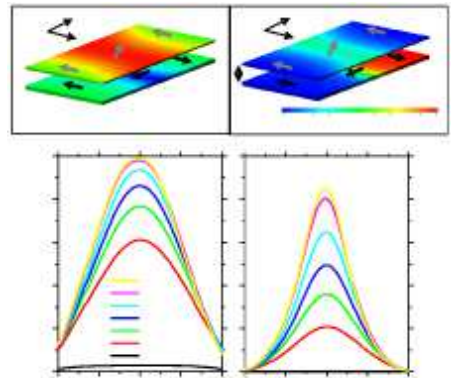}
\caption{Top: Simulations of a domain wall in the Co layer and its
influence on the FeNi magnetization, for a spacer thickness
\textit{d} of 30 nm.  In the top left panel, the component of the
magnetization along x (m$_x$) is shown color-coded for the Co
(bottom) and FeNi (top) layers, while in the top right panel the
y-component (m$_y$) is given. The rectangles represent only a small
fraction of the three-dimensional finite-element mesh, most of which
has been removed to uncover the region of interest. Bottom :
X-component (left) and y-component (right) of the FeNi magnetization
along x, in the middle of the simulated region, for different
distances \textit{d} between the Co and FeNi layers.}
\end{figure}

The effect of stray fields of domain walls in one layer on the
magnetization of the other layer in FM/NM/FM trilayers has been
treated quantitatively by several authors
\cite{fuller62,kuch03,schaf04,thomas00,lew03}. In most of these
cases, the domain wall was treated as a homogeneously magnetized
region of width W$_{DW}$ and with a magnetization perpendicular to
the overall magnetization direction. In order to get a more precise
idea of the influence of stray fields emitted by real domain walls,
we have performed micromagnetic simulations using a code based on a
combination of the finite element method (FEM) and the boundary
element method (BEM). This FEM/BEM scheme is particularly suited to
simulate magnetostatic interactions of ferromagnetic particles, as
described in Ref. \cite{hertel01}. The magnetic structures are
obtained by energy minimization. For the magnetic saturation
polarization J$_S$ = $\mu_0$M$_S$ and the exchange constant A,
values were taken of 1 T and 13 pJ/m for FeNi, and 1.76 T and 30
pJ/m for Co, respectively. Experimentally determined values of 1720
J/m$^3$ (FeNi) and 11200 J/m$^3$ (Co) were used for the uniaxial
anisotropy constant K. The magnetization was simulated in platelets
with a width of 200 nm along the easy magnetization axis (y-axis)
and 400 nm perpendicular to this axis (x-axis). To mimic extended
strips, free boundary conditions along \textit{y} were obtained
geometrically by connecting the edges of the platelet to form a
ribbon shape. Two domains with opposite magnetization directions
along the easy axis were introduced in the Co layer, resulting in a
N\'{e}el-type domain wall. The FeNi layer in the simulations was
initially homogeneously magnetized along –\textit{y}, and then its
magnetization was relaxed to reach equilibrium. The final
magnetization directions are indicated with grey and black arrows
for the FeNi and Co layers, respectively. The results of the
simulations for a 30 nm thick spacer layer are shown in the top
panels of Fig.~3. In the top left panel the \textit{x}-component of
the magnetization (m$_x$ = M$_x$/M$_S$, with M$_S$ the saturation
magnetization) is given, while the top right panel represents the
\textit{y}-component (m$_y$ = M$_y$/M$_S$). The color code ranges
from blue (m$_{x,y}$ = -1) to red (m$_{x,y}$ = 1) for all layers,
with m$_{x,y}$ = 0 given by green.

The simulated width of a domain wall in the Co layer using Lilley's
definition \cite{lilley50} is about 70 nm. The influence of the
domain wall on the magnetization of the FeNi layer decreases with
increasing separation between the FM layers but is considerable up
to spacer thicknesses as large as 100 nm.

The main result of the simulations is that above the Co domain wall
the magnetization in the FeNi layer is strongly tilted in the
direction perpendicular to the easy magnetization direction,
opposite to the magnetization direction in the center of the wall.
The profile of the \textit{x}-component of the FeNi magnetization
for different separations between the two layers is shown in the
bottom left panel of Fig.~3, while the \textit{y}-component is given
in the bottom right panel of Fig.~3. The \textit{x}-component of the
magnetization does not completely go to zero even at a distance of
200 nm of the center of the Co domain wall. The \textit{y}-profile
shows that the magnetization above the center of the Co DW is not
perfectly perpendicular to the easy axis but makes an angle
\textit{i}. This tilt angle \textit{i} depends on the separation
between the two layers, and a value of 81${^{\circ}}$ with respect
to the easy axis is found for a separation of 2.6 nm (the
experimental thickness of the alumina layer in our sample). Our
simulations show that the induced magnetization angle in the FeNi
layer also has an influence on the shape of the Co domain wall: it
becomes wider when the interlayer distance gets smaller.

The integrated width of the region with tilted FeNi magnetization,
taken from the simulated m$_y$ curve for a separation of 2.6 nm
between the layers, is about 150 nm. From the time-resolved magnetic
domain image in Fig.~2(g) we can get an experimental estimate of
this width. Taking into account the experimental resolution of
1~$\mu$m and the experimental contrast, we find a value of about 250
nm, which agrees reasonably well with the simulations. We can not
exclude, however, that more complicated structures, like
360${^{\circ}}$ domain walls, are formed above the Co domain walls
in the experiment.

In conclusion, the combination of nanosecond time-resolved XMCD-PEEM
measurements and micromagnetic simulations has allowed observing and
explaining the preferential nucleation of reversed FeNi domains
above Co domain walls in FeNi/Al$_2$O$_3$/Co trilayers. For magnetic
field pulses applied along the easy magnetization axis, the torque
on the FeNi moments is very small when these are aligned
anti-parallel to the applied field. Above the Co domain walls, the
FeNi moments are tilted away from this axis and the torque acting on
them is thus much larger. The Co domain wall stray field acts as a
transverse bias field that locally decreases the energy barrier for
nucleation, significantly increasing the local switching speed. The
same principle has been exploited to obtain ultrafast magnetization
switching in small magnetic structures, using an external transverse
bias field \cite{choi01} This finding is of importance also for
current-induced domain wall propagation in trilayer systems, a
subject that is widely investigated nowadays because of the
potential use in MRAM systems. The effect of domain wall stray
fields can be controlled by manipulating the width and position of
the domain wall. This allows increasing the local speed and
reproducibility of magnetic switching.

We thank A. Vaur\'{e}s for sample preparation. Financial support by
BMBF (no.~05, KS1EFA6), EU (BESSY-EC-HPRI Contract No.
HPRI-1999-CT-00028) and the Laboratoire Europ\'{e}en Associ\'{e}
'Mesomag' is gratefully acknowledged. J.C. acknowledges support
through a "Ram\'{o}n y Cajal" contract and through project No.
MAT2003-08627-C02-02 from the Spanish Ministry of Science and
Technology.


\begin{thebibliography}:
\bibitem{fuller62} H.W. Fuller and D. L. Sullivan, J.~Appl.~Phys. \textbf{33}, 1063 (1962).

\bibitem{Hubertbook} A.~Hubert and R.~Sch\"{a}fer: \textit{Magnetic Domains:
The Analysis of Magnetic Microstructures} (Springer-Verlag, Berlin,
1998),p.488ff.

\bibitem{kuch03} W.~Kuch, L.I.~Chelaru, K.~Fukumoto, F.~Porrati, F.~Offi, M.~Kotsugi, and J.~Kirschner,
Phys.~Rev.~B \textbf{67}, 214403 (2003).


\bibitem{schaf02} R.~Sch\"{a}fer \textit{et al.},
Phys.~Rev.~B \textbf{65}, 144405 (2002).

\bibitem{schaf04} V.~Christoph and R.~Sch\"{a}fer, Phys.~Rev.~B \textbf{70}, 214419 (2004).


\bibitem{Wiebel05} S.~Wiebel \textit{et al.}, Appl. Phys. Lett.
\textbf{86}, 142502 (2005).

\bibitem{thomas00} L.~Thomas, M.G.~Samant, and S.S.P.~Parkin, Phys.~Rev.~Lett. \textbf{84}, 1816 (2000).

\bibitem{choi01} B.C.~Choi, M.~Belov, W.K.~Hiebert, G.E.~Ballentine, and
M.R.~Freeman, Phys.~Rev.~Lett. \textbf{86}, 728 (2001).



\bibitem{Sussiau} M.~Sussiau \textit{et al.}, J.~Magn.~Magn.~Mater. {\bf 165}, 1 (1996).


\bibitem{Montaigne} F.~Montaigne \textit{et al.}, {Appl.~Phys.~Lett.} {\bf 76}, 3286 (2000).


\bibitem{pennec04} Y.~Pennec \textit{et al.}, Phys.~Rev.~B \textbf{69}, 180402(R) (2004).

\bibitem{Neel62} L.~N\'{e}el, {C.R.~Acad.~Sci.~Paris} \textbf{255}, 1676
(1962).

\bibitem{encinas03} A.~Encinas-Oropesa and F.~Nguyen~Van~Dau,
J.~Magn.~Magn.~Mater. \textbf{256}, 301 (2003).


\bibitem{vogel03} J.~Vogel \textit{et al.}, Appl.~Phys.~Lett.
\textbf{83}, 2299 (2003); J.~Appl.~Phys. \textbf{83}, 2299 (2004).


\bibitem{kuch04} W.~Kuch \textit{et al.}, Appl.~Phys.~Lett. \textbf{85}, 440 (2004);
J.~Vogel \textit{et al.}, Phys.~Rev.~B \textbf{71}, 060404(R)
(2005).


\bibitem{schneider04} C.M.~Schneider \textit{et al.}, Appl.~Phys.~Lett. \textbf{85},
2562 (2004); S.B.~Choe \textit{et al.}, Science \textbf{304}, 420
(2004).


\bibitem{kuch98} W.~Kuch \textit{et al.}, Surf.~Rev.~Lett. {\bf 5}, 1241 (1998).

\bibitem{lew03} W.S.~Lew, S.P.~Li, L.~Lopez-Diaz, D.C.~Hatton, and
J.A.C.~Bland, Phys.~Rev.~Lett. \textbf{90}, 217201 (2003).


\bibitem{hertel01} R.~Hertel, J.~Appl.~Phys. \textbf{80}, 5752
(2001).

\bibitem{lilley50} B.A.~Lilley, Phil.~Mag. \textbf{41}, 792 (1950).

\end{thebibliography}
\end{document}